\documentclass[%
preprint,
amsmath,amssymb,
aps,
prb, showkeys,
]{revtex4-2}

\usepackage{graphicx}
\usepackage{dcolumn}
\usepackage{bm}
\usepackage[mathlines]{lineno}
\linenumbers\relax 

\usepackage{natbib}

\newcommand{\BSTO}{Ba$_{1-x}$Sr$_x$TiO$_3$}
\newcommand{\BTO}{BaTiO$_3$}
\newcommand{\STO}{SrTiO$_3$}
\newcommand{\LSNO}{La$_{1.1}$Sr$_{0.9}$NiO$_3$}
\newcommand{\pola}{$^{+}$}
\newcommand\ion[2]{${}^{#1}$#2\pola}

\newcommand{\GEMAC}{Université Paris-Saclay, UVSQ, CNRS, GEMaC, 78000, Versailles, France}
\newcommand{\GREMAN}{GREMAN UMR 7347, Université de Tours, CNRS, INSA-CVL, 16 rue Pierre et Marie Curie, 37071 Tours, France}
\newcommand{\Suzhou}{School of Mathematics and Physics, Suzhou University of Science and Technology, Suzhou 215009, China}

\begin{document}

\nolinenumbers 

\title{Evidence of the matrix effect on a compositionally graded oxide thin film}

\author{J. Scola}
\affiliation{\GEMAC}
\author{F. Jomard}
\affiliation{\GEMAC}
\author{E. Loire}
\affiliation{\GEMAC}
\author{J. Wolfman}
\affiliation{\GREMAN}
\author{B. Negulescu}
\affiliation{\GREMAN}
\author{G. Z. Liu}
\affiliation{\Suzhou}
\author{M.-A. Pinault-Thaury}
\affiliation{\GEMAC}


\begin{abstract}
A heterostructure of \BSTO /\LSNO /\STO{} has been analysed by magnetic sector secondary ion mass spectrometry (SIMS).
The stoichiometry parameter $x$ of the top layer was made varying continuously from 0 to 1 along the width of the sample by combinatorial pulsed laser deposition.
Prior to SIMS analysis, the composition gradient of \BSTO{} was quantitatively characterized by chemical characterizations including wavelength and energy dispersive X-ray spectroscopies.
Even if the Ti content is constant into \BSTO{}, its ionic yield exhibits an increasing trend as Ba is substituted by Sr.
Such a phenomenon can be explained by the variation of the neighbouring atoms chemistry which affects the ionization probability of titanium during the sputtering process.
In addition to the continuously varying composition, the oxide multilayer sample features sharp interfaces hence the in-depth resolution under our analysing conditions has been investigated too.
The modelling of the interface crossing profiles reveals that the instrumental contribution to the profile broadening is as low as 5 nm.
\end{abstract}


\maketitle

\section{Introduction}
Due to their high dielectric permittivity, low dielectric losses and dielectric tunability \cite{qiuStructureDielectricCharacteristics2016a}, dielectric materials are promising for applications in microwave and multifunctional devices \cite{baoBariumStrontiumTitanate2008, jinNovelMultifunctionalProperties2009}.
\BSTO{} is one of the most famous dielectrics since the Sr fraction can be varied, thus tailoring its dielectric properties. So far, attention has been paid to its fabrication with different Ba/Sr ratios in series of samples having a given composition.
Among the methods developed for thin film growth, combinatorial pulsed laser deposition (PLD) provides a unique chance to prepare continuous composition spread libraries in the plane of a single substrate at the same time \cite{wolfmanInterfaceCombinatorialPulsed2020}.
Then, the comparison of the properties of the various compounds solely depends on the local chemical composition of the sample, contrary to a batch of  samples influenced by run-to-run fabrication parameter alterations that cannot be ruled out. 

This original growth technique offers an ideal ground for exploring the secondary ion mass spectrometry SIMS matrix effect.
This effect designates variations of signal which are not related to the atomic concentration.
The secondary ion intensity collected by the instrument indeed combines several factors including the  concentration of the atom of interest in the bombarded material, its the sputtering yield and its ionization efficiency.
The latter two depends on the chemical and morphological properties of the matrix.
Therefore significant variations of the ionic intensity can occur from sample to sample independently to the atomic concentration which is further sought.
Lots of efforts have been made to elucidate the effect of a matrix on the ionic yield of diluted traces.
Although empirical trends have been identified very early \cite{andersenThermodynamicApproachQuantitative1973}, the vast variety of the material chemistry and of the analysis conditions have made the task tedious and little success to build a unified theory for all cases have been met so far \cite{williamsSputteringProcessSputtered1979b}.
A common approach to reach SIMS quantitative measurements consists of interpolating among chemically similar standards using semi-empirical models with limited validity domains \cite{eilerSIMSAnalysisOxygen1997}.
This means to track ion yields in samples with similar compositions; it can be either series of alloys \cite{priebeMatrixEffectTOFSIMS2020} 
or multi-implanted samples \cite{prudonSIMSQuantificationThick2013}.

Here, we report the benefits of the combinatorial chemistry to undoubtedly evidence a matrix effect in a complex oxide.
We also took advantage of the sharpness of our sample interfaces to determine the depth resolution in our analysis conditions.


\section{Experiment}
\subsection{Sample}
\begin{figure}
\includegraphics[width=.35\textwidth]{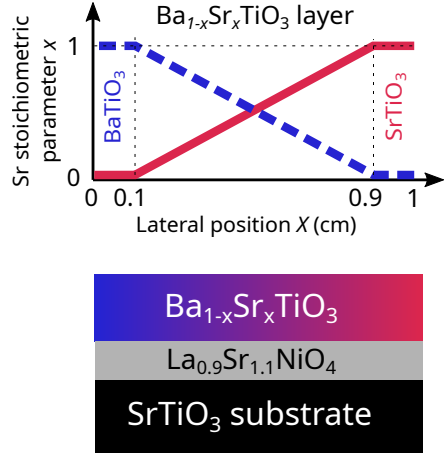}
\caption{
\label{fig:GZ46_sketch}
Upper panel: scheme of the Ba and Sr content along the compositional gradient in the \BSTO{} layer;
lower panel: sketch of the multilayer.
}
\end{figure}

The studied sample consists in a two oxide layers grown on a commercial SrTiO$_3$ substrate by combinatorial PLD. 
The deposition conditions detailed in Ref.  \cite{sandeepEvaluationOpticalAcoustical2021a}.
The intermediate layer of \LSNO{} acts as a bottom electrode.
The top layer is a so-called \emph{material library} that is a compositionally graded film ranging from BaTiO$_3$ on one side to SrTiO$_3$ on the other side. 
It includes in-between every intermediate compounds \BSTO{} with $0\le x \le 1$ (Fig. \ref{fig:GZ46_sketch}).
The combinatorial chemistry enables to test a large number of compositions without 
any growth-to-growth discrepancies.
In addition, all compositions are deposit on the same substrate avoiding the growth difficulties of stack of layers and the substitutional approach preserves the crystallographic structure better than implantation.

The local chemistry of the sample was controlled by wavelength dispersive X-ray spectroscopy (WDS) and energy dispersive X-ray spectroscopy (EDS): the stoichiometric parameter of strontium increases linearly from $x=0$ to $x = 1$ as it substitutes barium along the gradient axis from $X = 0$ to $X = 1$ cm.
The quality of the growth was evidenced by X-ray diffraction showing that both out-of-plane and in-plane lattice parameters vary linearly with the strontium content in coherence with the reference compounds \cite{sandeepEvaluationOpticalAcoustical2021a, qiuStructureDielectricCharacteristics2016a}.


\begin{figure}[htp!]
\begin{center}
\includegraphics[width=.8\textwidth]{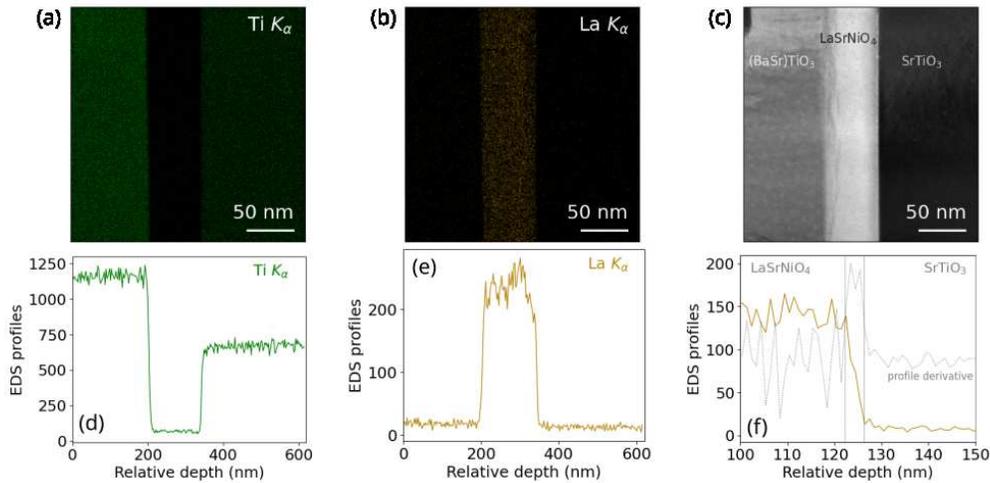}
\caption{
\label{fig:EDS}
EDS map at the $K_\alpha$ edges of titanium (a) and lanthanum (b) and their integrated profiles across the depth of the sample (d and e respectively).
High-angle annular dark-field image of the mapped area labelled with the layer compositions.
Detail of the \LSNO/\STO{} interface integrated profile of the lanthanum distribution mapped at a greater magnification (f).
}
\end{center}
\end{figure}

The chemical composition was further investigated by electron dispersive spectroscopy (EDS) analysis on a transmission electron microscope (TEM).
The EDS chemical mapping was performed on a cross-section of the sample.
Figure \ref{fig:EDS} reports the EDS maps for titanium (a) and lanthanum (b) as well as their integrated profiles  (Fig. \ref{fig:EDS}-d,e).
The three compounds of the sample are labelled on the corresponding dark field image (Fig. \ref{fig:EDS}-c).
The maps and integrated profiles evidence a sharp interface between the \LSNO{} layer and the \STO{} substrate.
At greater magnification, the inter-diffusion length at the interface is estimated from the half-width at the half-maximum of the profile derivative: it is less than 2 nm, close to the experimental resolution.

\subsection{SIMS conditions}

The SIMS analyses were performed with a magnetic sector analyser (IMS7f-CAMECA).
To avoid surface charging, the sample surface was coated with gold (50 nm) prior analyses. We employed the Cs$^+$/$M^+$ configuration: positive primary ions with a Cs$^+$ source and detection of positive secondary $M^+$ ions of mass $m$.
The energy of the Cs$^+$ primary beam is set to 10 keV.
The raster size is 150$\times$150 $\mu$m$^2$ with an analyzed zone restricted to a diameter of 33 $\mu$m and an electronic gate set to 70\%.
The analysis chamber is kept under high vacuum ($\approx$ 10$^9$ mbar).
Secondary ions are detected in the positive mode by biasing the sample to +5 kV, leading to an interaction energy of the primary ions of 5 keV and an incidence angle of 46 $^\circ$ with respect to the normal of the sample.
We used a primary intensity of 10 nA.
The depth of the resulting SIMS crater is measured with a Dektak8 step-meter. Knowing the total sputtering time for each oxide layer, we measured the sputtering rate in each of them.
The values found for the \BSTO{} layer, the \LSNO{} layer and the \STO{} substrate were of 0.25, 0.36, 0.24 nm/s respectively.
Within the experimental resolution, no dependence of the sputtering rate was observed with the position in the compositionally graded layer.
This excludes any large variations of the sputtering process, including the Cs$^+$ primary ion implantation regime related to the changes in the \BSTO{} composition.

The optimization of the analysis conditions is made challenging by the complexity of the sample. The elements in presence spread over wide ranges of masses and ionic intensities while the layers are thin. Detection of low intensities without increasing too much the sputtering rate requires a compromise on the mass resolution.
We performed SIMS analyses at low mass resolution, $m/\Delta m \approx 400$.
As a result, we had a maximum sensitivity and the time for cycling masses was made short enough to get a hundred of measurements within each layer.
In so doing, the number of points that describes the interfaces is enough and does not require to slow down the sputtering rate, which might have pulled the weakest signals below the detection limit.

\begin{figure}[ht!]
\includegraphics[width=.8\textwidth]{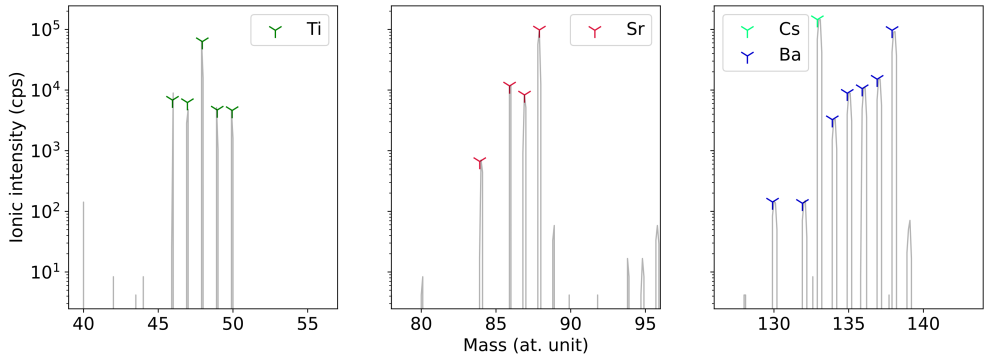}
\caption{
\label{fig:MassSpectra}
Experimental mass spectra (solid line) around the masses of titanium, strontium and barium collected in the \BSTO{} layer.
The markers denote the natural isotopic abundances.
}
\end{figure}

Due to the complexity of the sample layers, a particular care was given to the possible mass interferences. Indeed, the extraction energy was offset by -60 V in order to reduce the contribution of molecular ions as their intensities tend to decay faster in energy than the single ions \cite{stevie2015secondary}. 
Moreover, we verified that the mass spectra acquired in the compositionally graded layer actually match the natural relative abundance [13] of the selected elements (Figure \ref{fig:MassSpectra}).
For example, the relative ionic intensities collected at integer masses from 46 to 50 corresponds to the natural abundance of \ion{46}{Ti}, \ion{47}{Ti}, \ion{48}{Ti}, \ion{49}{Ti} and \ion{50}{Ti} which are 8.25\%, 7.44\%, 73.72\%, 5.41\% and 5.18\% respectively \cite{kondevNUBASE2020EvaluationNuclear2021}.
This ensures that the contribution of single ions to the intensities collected at the analysed masses are significantly larger than the one of interfering molecular ions, if any. 
To minimize the complexity of the mass interferences, the analysed masses were chosen as small as possible and focused on the single ions \ion{46}{Ti}, \ion{48}{Ti}, \ion{86}{Sr}, \ion{88}{Sr}, \ion{138}{Ba} and \ion{139}{La}.

\section{Results and discussion}
\subsection{Matrix effect}
\begin{figure}
\includegraphics[width=.5\textwidth]{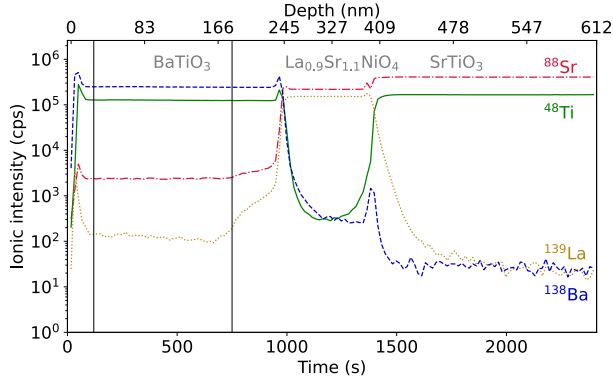}
\caption{
\label{fig:DepthProfile}
Ionic intensities of \ion{88}{Sr} (dash dotted line), \ion{48}{Ti} (solid line), \ion{139}{La} (dotted line) and \ion{138}{Ba} (dashed line),  plotted as a function of the sputtering time (lower axis) and the sample depth  calculated from the sputtering rates measured for each layer. 
The vertical lines denotes the range of data points where the \ion{48}{Ti}, \ion{88}{Sr} and \ion{138}{Ba} average intensities were calculated
}
\end{figure}

The depth profiles of the sample in the Ba-rich side of the composition gradient are reported in Figure \ref{fig:DepthProfile} where the ionic intensities are plotted as a function of the sputtering time (lower x-axis) and sputtering depth (upper x-axis).
All selected elements have ionic yields above $10^5$ cps.
The \ion{138}{Ba} intensity steadily remains high in the top layer and decays in the rest of the structure.
Conversely, the \ion{88}{Sr} intensity rapidly rises at the interface between \BSTO{} ($x \approx 0.1$) and \LSNO{} where it reaches a plateau value and persists at a high level in the \STO{} substrate.
The intermediate nickelate layer is denoted by the \ion{139}{La} intensity plateau.
It should be mentioned that the sharp signal variations occurring at the interfaces come from the non-stationarity of the sputtering regime as interfaces are crossed;
for each analysis position, the intensity of each ion in \BSTO{} is estimated by the average of at least a hundred data points taken in the plateaus, away from the surface and the interface with \LSNO{} (in the region marked out by the vertical lines in Fig. \ref{fig:DepthProfile}).

\begin{figure}
\includegraphics[width=.5\textwidth]{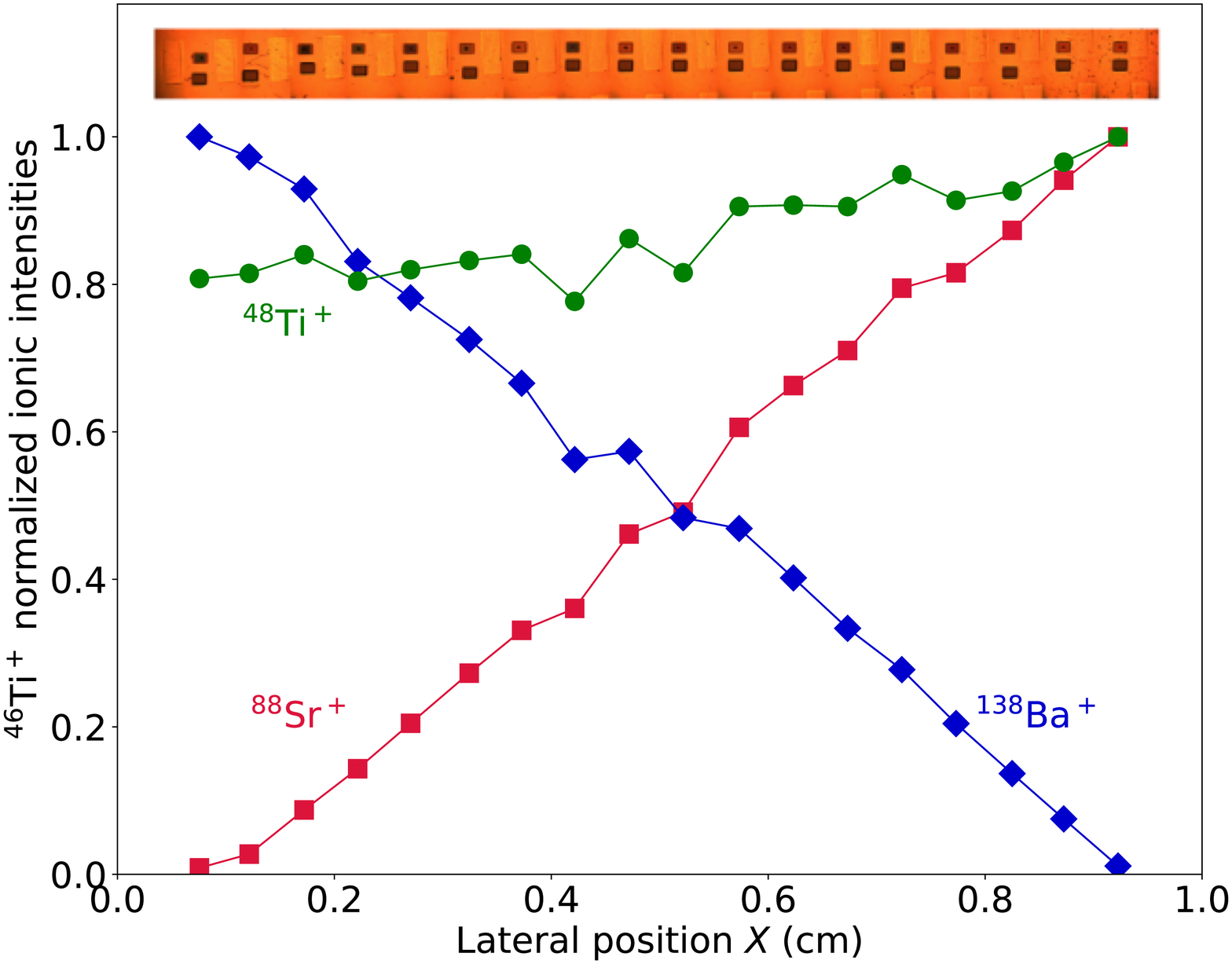}
\caption{
\label{fig:matrix_effect_BaSrTi}
Ionic intensities of \ion{48}{Ti} (green bullets), \ion{88}{Sr} (red squares), \ion{138}{Ba} (blue squares) measured in the \BSTO{} layers as well as \ion{48}{Ti} measured in the \STO{} substrate (green and black bullets) normalized by \ion{46}{Ti} intensity measured in the \STO{} substrate.
The ranges between the background levels to the maximum values are rescaled from zero to unity.
Inset: stitching of optical images taken from the ISM7f camera of the analyses craters corresponding to the data points.
}
\end{figure}

Depth profiles for the selected masses were collected under the same conditions at evenly spaced positions along the chemical composition gradient; 
a stitching of optical images of the analyses craters is inserted in the inset of Figure \ref{fig:matrix_effect_BaSrTi}.
One of the greatest interests of the sample is the presence of the \STO{} substrate underneath compositionally graded layer.
Its composition is close or identical to that of the \BSTO{} top layer so it can be used as a constant reference along the gradient direction.
Moreover, each reference is acquired in the same depth profile as the signal from the layer of interest.
This makes the normalization procedure very robust and sound.
As expected, no dependence of the ionic intensities from the substrate with the lateral position $X$ is observed in the substrate. 
The small variations are typical of irreproducibility of the instrument collecting factor as the lateral position is changed.
The isotopic ratios of both titanium and strontium fit the natural abundance, thus confirming the absence of mass interferences at the analysed masses in the substrate too.
In the following, those substrate \ion{46}{Ti} intensities will be used to normalize the intensities measured in the graded film.
%
The normalized intensities of \ion{88}{Sr} and \ion{138}{Ba} measured in the \BSTO{} layer are plotted against the $X$ position in Figure \ref{fig:matrix_effect_BaSrTi}.
The opposite trends of these two elements are in very good agreement with the chemical gradient as it was evidenced Ref. \cite{sandeepEvaluationOpticalAcoustical2021a}.
The intensities of Sr and Ba vary linearly with the Sr stoichiometric parameter $x$. 

The situation for titanium is quite different.
As Ba is substituted by Sr, the \ion{48}{Ti} signal level increases by about 25 \% in the \BSTO{} graded layer while it concentration remains constant.
This evidences a well-characterised matrix effect that is
a significant dependence of the titanium ionic signal on its chemical environment.

In our case, any growth-to-growth uncontrolled discrepancies can be excluded. Indeed, a compositionally graded layer makes it possible to sample many chemical compositions while strictly preserving the content of matrix elements together with the crystallographic structure and orientation, known to intervene in the sputtering process.

\begin{figure}
\includegraphics[width=.5\textwidth]{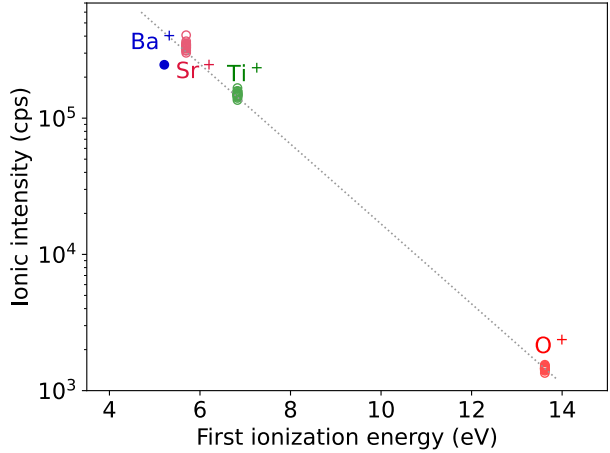}
\caption{
\label{fig:intensity_vs_Eion}
Aggregation of the intensities of \ion{88}{Sr}, \ion{48}{Ti} and \ion{16}{O} collected at any positions in the \STO{} substrate along with the intensity of \ion{138}{Ba} collected in the \BSTO{} layer at $x = 0$ plotted against the first ionization energy of the corresponding atoms \cite {Ralchenko2000NISTAS}.
The dotted line is a guide for the eyes.
}
\end{figure}

The remaining changes of composition as Ba is substituted by Sr is likely to affect the \ion{48}{Ti} intensity in three ways :
(i) by modifying the electronegativity of the neighboring atoms of Ti and in turn changing its ionization probability,
(ii) by changing  the average atomic mass involved in the collision chains triggered by the primary ion bombardment or
(iii) a combination of both effects.
Interestingly, the logarithm of the ionic intensities of the four elements composing the \BSTO{} layer turns out to scale with their atomic ionization energy (Fig. \ref{fig:intensity_vs_Eion})
disregarding other parameters like their mass.
This is coherent with empirical trend observed in pure metals and semiconductors \cite{williamsUniversalModelSputtered1980a} and it suggests that in such complex oxides first ionization energy remains a major parameter for the ionization process in general.
For the particular case of the \ion{48}{Ti} formation it would mean that the more energy it takes to ionize the neighbouring atoms, the more probable the formation of a Ti$^+$ ion is.
A contribution of the sputtering process to the matrix effect is not excluded but made little probable by the fact that the variations of the sputtering rates as a function of the \BSTO{} composition are very small.

\subsection{Depth resolution}
Beside matrix effect, the sample can be used as a reference for another SIMS specific response.
Investigating how its sharp interfaces are crossed during the analysis provides an efficient assessment of the instrumental depth resolution in complex oxides under the chosen set of analysis conditions.
The \LSNO / \STO{} interface is crossed by each of the eighteen depth profiles reported before but not affected by the compositional gradient of the top layer.
Among the selected ions, \ion{139}{La} has profiles that gather many advantages:
more than two decades of dynamic range, no mass interference and small signal exaltation due to the non-stationary sputtering regime at the interface.

In the framework of the atomic mixing-roughness-information depth (MRI) depth model, the depth profile of an ideally sharp interface is broadened by three mechanisms \cite{hofmannDepthResolutionDepth1999}.
First, the sputtering process induces the mixing of the atoms within a region that extends over a characteristic length $w$; 
since the original atomic order is lost in the atomic mix, its size sets the ultimate depth resolution of the analysis.
From the numerical point of view, the Heaviside step function modelling an ideally sharp interface gets convoluted by 
\begin{equation}
g_w (z-z^\prime) = \frac{1}{w} \exp\left[-(z-z^\prime + w)/w \right], \quad z^\prime - w < z.
\end{equation}
The exponential function introduces an asymmetric shape related to the fact that the ionic emission is delayed by the atom retention in the mix.

Second, the interface roughness contributes also to the width of the depth profile.
Indeed, it is responsible for the presence in a given layer of atoms originating from another layer beyond the nominal (\textit{i.e.} ideally sharp) interface position and vice versa.
The interface roughness can originate from the surface roughness that is transferred to the bottom of the analysis crater by the sputtering process.
In addition, inter-diffusion can lead to an equivalent situation.
Whatever the nature of the roughness $\sigma$ is, the associated convolution function is Gaussian:
\begin{equation}
g_\sigma (z - z^\prime) = \frac{1}{\sigma \sqrt{2\pi}} \exp \left[ -(z - z^\prime)^2 / (2\sigma^2)\right].
\end{equation}

The third mechanism is related to the information depth.
It is not relevant here as the collected information is expected to be purely superficial in SIMS analysis.

In practise, the modelling of a depth profile across a real interface consists in considering  the experimental profile as a Heaviside function $H(z)$ and then calculating its convolution product by $g = g_w \otimes g_\sigma$ using the set of parameters $(w, \sigma)$ that fits best the experimental  profile:
\begin{equation}
\label{eq:MRI}
I_{MRI} (z) = \int_{-\infty}^\infty H (z^\prime) g (z-z^\prime) \mathrm{d}z^\prime.
\end{equation}
Here, $H(z)$ switches from the constant intensity of \ion{139}{La} in \LSNO{} to the background noise at $z = 0$ as lanthanum is not present in the \STO{} substrate.

\begin{figure}
\begin{center}
\includegraphics[width=.5\textwidth]{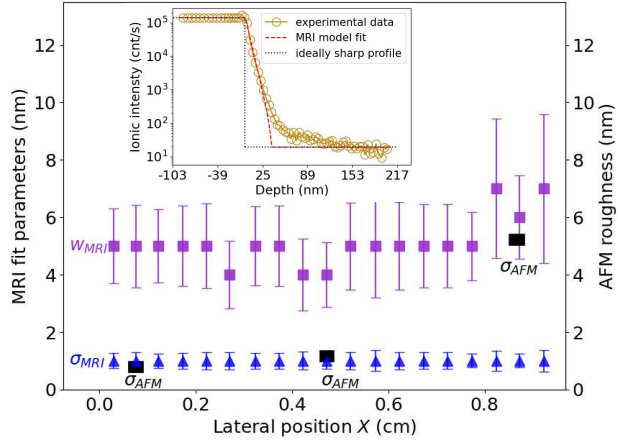}
\caption{
\label{fig:MRI_analysis}
Best fit values of MRI atomic mix extension $w$ and interface roughness $\sigma$ reported for all analysed $X$ positions.
Error bars are calculated from the standard deviation of the fit values with respect to the data points: $\varepsilon = \left( \frac{1}{N}\sum \lvert {I_{exp} - I_{MRI}}\rvert^2 \right)^{1/2} $.
Black squares represents the roughness measured by AFM at the bottom of sputtering craters. 
Inset: Depth profile of \ion{139}{La} (yellow circles) around the \LSNO{}/\STO{} interface (set at the origin). The dotted line represents the ideally sharp profile (Heaviside function). The red dashed line represents the MRI model calculated using Eq. (\ref{eq:MRI}) with the best set of fitting parameters $w = 5$ nm and $\sigma = 1$ nm (red dashed line).
}
\end{center}
\end{figure}

The MRI modelling the depth profiles at the \LSNO /\STO{} interface was performed for each $X$ position on the sample.
The best fit parameters ($w$, $\sigma$) and the fit errors are reported in Fig. \ref{fig:MRI_analysis}.
The inset of Figure \ref{fig:MRI_analysis} shows a typical depth profile of \ion{139}{La} at the \LSNO / \STO{} interface and its MRI modelling using the best fitting parameters.
The agreement of the fit with the experimental data is very good, especially for the highest intensities, away from the background noise.
The same agreement was found for all analysed $X$ positions.
This indicates that the atomic mix and interface roughness modelled by Eq. \ref{eq:MRI} are coherent with the actual broadening of the measured depth profiles.
The values of $w$ and $\sigma$ consistently stand close to 5 nm and 1 nm respectively.
We consider such a repeatability as a statistical confirmation of the model validity since no dependence on the top layer gradient is expected due to the fact that the \LSNO{} layer was uniformly grown on the \STO{} substrate.
The exception of three points whose $w$ parameter values deviate from the main trend can be related to the sample itself as discussed in the following.

The value of $\sigma$ has to be compared to the interface roughness which was measured by atomic force microscopy (AFM, Bruker Icon) at the bottom of dedicated sputtering craters interrupted right at the \LSNO/\STO{} interface.
Topographic images were acquired at three positions of sample corresponding to the barium-rich, intermediate composition and strontium-rich regions in the compositional gradient. 
The estimated rms roughness is 0.8, 1.18 and  5.24 nm respectively.
The relatively large rms roughness reported at the bottom of the crater sputtered in the \STO{} side of the gradient coincides with the deviation of the $w$ fit parameter in the Sr rich region.
This suggests that, at one end of the gradient, the interface quality of the sample may actually depart from its best sharpness, thus affecting the relevance of the MRI mode.
The roughness in the rest of the sample turns out to be smaller than the inter-diffusion length at the interfaces estimated by the EDS mappings.
The very good agreement of the measured roughness and its counterpart $\sigma$  output by the MRI model further supports the hypothesis that the model is valid in this context.
Then, the fit parameter $w$ can be considered as a realistic estimation of the actual atomic mixing extent. 
This means that a small fraction of the experimental depth profile width comes from the actual interface roughness and most of it is due to the instrumental convolution.
The consequent depth resolution is as small as for analyses using ten times lower primary beam energies. \cite{morrisO2ProbesampleConditions2013}.

\section{Conclusion}
We reported SIMS analysis on an oxide sample consisting in two oxide layers grown on a commercial \STO{} substrate.
The top layer is a dielectric material library of 
\BSTO{}, that is a continuous composition gradient from \BTO{} to \STO{} along the width of the sample, the Ti content being unchanged.
The Ba and Sr ionic intensities turned out to linearly depend on the concentration.
On the other hand, the concentration of Ti is the same for all compositions but its ionic intensity increases by 25 \% as Ba is substituted by Sr.
This evidences a matrix effect that can be associated to the slight decrease of ionisation energy of the environment from the Ba-rich region to the Sr-rich region.
The generality of this explanation would require some further investigations in other sets of compounds.
Besides, we took advantage of the sharp interfaces between the layers to evaluate the depth resolution in the chosen SIMS conditions.
The atomic mixing-roughness-information depth model was shown to yield a realistic estimation of the volume covered by the atomic mixing region in our analysis conditions.
The corresponding instrumental depth resolution is of the order of 5 nm which is suitable for investigations of oxide thin films.

\section{Acknowledgements}
This work was financially supported by Agence National de la Recherche  via the project SUPERNICKEL
(ANR-21-CE30-0041).
The FIB experiments were supported by the French RENATECH network, the CPER Hauts de France project IMITECH and the Métropole Européenne de Lille.
We acknowledge the Institute of Mineralogy, Physics of Materials and Cosmochemistry (IMPMC UMR 7590 CNRS Sorbonne Université) for the access of the transmission electron microscope (TEM) Jeol 2100F and Prof. Nicolas Menguy for his help in data acquisition. The TEM facilities at IMPMC were purchased owing to a support by Region Ile-de-France grant SESAME 2000E1435.
EDS data were processed using HyperSpy python package \cite{francisco_de_la_pena_2023_10151686}.

\bibliography{sims24}{}

\bibliographystyle{apsrev4-2}

\end{document}